\documentclass[superscriptaddress,preprintnumbers,aps,prb,twocolumn,10pt,longbibliography]{revtex4-2}

\usepackage{amsfonts}
\usepackage{amssymb}
\usepackage{amsmath}
\usepackage{graphicx}
\usepackage{float}
\usepackage{multirow}

\usepackage[usenames,dvipsnames]{color}

\usepackage[hidelinks]{hyperref} 

\begin{document}

\newcommand{\ie}{{i.e.,}}
\newcommand{\eg}{{e.g.,}}
\newcommand{\etal}{{\it et al.}}

\newcommand{\Kxx}{$\kappa_{\mathrm{xx}}$}
\newcommand{\Kxy}{$\kappa_{\mathrm{xy}}$}
\newcommand{\Kzy}{$\kappa_{\mathrm{zy}}$}
\newcommand{\Kzz}{$\kappa_{\mathrm{zz}}$}

\newcommand{\ncco}{Nd$_{2-x}$Ce$_x$CuO$_4$}
\newcommand{\pcco}{Pr$_{2-x}$Ce$_x$CuO$_4$}
\newcommand{\ndlsco}{La$_{1.6-x}$Nd$_{0.4}$Sr$_x$CuO$_4$}

\newcommand{\TN}{$T_{\mathrm{N}}$}
\newcommand{\Tc}{$T_{\mathrm{c}}$}

\newcommand{\xstar}{$x^{\star}$}

\title{Thermal Hall conductivity of electron-doped cuprates: Electrons and phonons}

\author{Marie-Eve~Boulanger}
\thanks{M.-E.~B. and L.~C. contributed equally to this work.}
\affiliation{Institut quantique, D\'epartement de physique \& RQMP, Universit\'e de Sherbrooke, Sherbrooke, Qu\'ebec, Canada}

\author{Lu~Chen}
\thanks{M.-E.~B. and L.~C. contributed equally to this work.}
\affiliation{Institut quantique, D\'epartement de physique \& RQMP, Universit\'e de Sherbrooke, Sherbrooke, Qu\'ebec, Canada}

\author{Vincent~Oliviero}
\affiliation{LNCMI-EMFL, CNRS UPR3228, Univ. Grenoble Alpes, Univ. Toulouse, INSA-T, Grenoble and Toulouse, France}

\author{David~Vignolles}
\affiliation{LNCMI-EMFL, CNRS UPR3228, Univ. Grenoble Alpes, Univ. Toulouse, INSA-T, Grenoble and Toulouse, France}

\author{Ga\"el~Grissonnanche}
\affiliation{Institut quantique, D\'epartement de physique \& RQMP, Universit\'e de Sherbrooke, Sherbrooke, Qu\'ebec, Canada}
\affiliation{Laboratoire des Solides Irradi\'{e}s, CEA/DRF/IRAMIS, CNRS, \'{E}cole Polytechnique, Institut Polytechnique de Paris, 91128 Palaiseau, France}
\affiliation{IRL Fronti\`{e}res Quantiques, Universit\'{e} de Sherbrooke--CNRS, Sherbrooke, Qu\'{e}bec, Canada}

\author{Kejun~Xu}
\affiliation{Geballe Laboratory for Advanced Materials, Stanford University, Stanford, California, USA}
\affiliation{Stanford Institute for Materials and Energy Sciences, SLAC National Accelerator Laboratory, Menlo Park, California, USA}
\affiliation{Departments of Physics and Applied Physics, Stanford University, Stanford, California, USA}

\author{Zhi-Xun~Shen}
\affiliation{Geballe Laboratory for Advanced Materials, Stanford University, Stanford, California, USA}
\affiliation{Stanford Institute for Materials and Energy Sciences, SLAC National Accelerator Laboratory, Menlo Park, California, USA}
\affiliation{Departments of Physics and Applied Physics, Stanford University, Stanford, California, USA}

\author{Cyril~Proust}
\affiliation{LNCMI-EMFL, CNRS UPR3228, Univ. Grenoble Alpes, Univ. Toulouse, INSA-T, Grenoble and Toulouse, France}
\affiliation{IRL Fronti\`{e}res Quantiques, Universit\'{e} de Sherbrooke--CNRS, Sherbrooke, Qu\'{e}bec, Canada}

\author{Jordan~Baglo}
\email[E-mail: ]{jordan.baglo@usherbrooke.ca}
\affiliation{Institut quantique, D\'epartement de physique \& RQMP, Universit\'e de Sherbrooke, Sherbrooke, Qu\'ebec, Canada}

\author{Louis~Taillefer}
\email[E-mail: ]{louis.taillefer@usherbrooke.ca}
\affiliation{Institut quantique, D\'epartement de physique \& RQMP, Universit\'e de Sherbrooke, Sherbrooke, Qu\'ebec, Canada}
\affiliation{IRL Fronti\`{e}res Quantiques, Universit\'{e} de Sherbrooke--CNRS, Sherbrooke, Qu\'{e}bec, Canada}
\affiliation{Canadian Institute for Advanced Research, Toronto, Ontario, Canada}

\date{\today}

\begin{abstract}
It has recently become clear that phonons generate a sizable thermal Hall effect in cuprates, whether they
are undoped, electron-doped or hole-doped (inside the pseudogap phase).
At higher doping, where cuprates are reasonably good metals, mobile electrons also generate a thermal Hall
effect, the thermal equivalent of the standard electrical Hall effect.
Here we show that in the cleanest crystals of the electron-doped cuprate \ncco, at high doping, the phonon
and electron contributions to the thermal Hall conductivity \Kxy~are of comparable magnitude, but of
opposite sign.
In samples of lower quality, phonons dominate \Kxy, resulting in a negative \Kxy~at all temperatures.
The fact that the negative phononic \Kxy~in the metallic state is similar in magnitude and temperature
dependence to that found in the insulating state at lower doping rules out any mechanism based on
skew scattering of phonons off charged impurities, since a local charge should be screened in the
metallic regime. 
The phononic \Kxy~is found to persist over the entire doping range where antiferromagnetic correlations
are known to be significant, suggesting that such correlations may play a role in generating the
phonon thermal Hall effect in electron-doped cuprates.
If the same mechanism is also at play in hole-doped cuprates, the presence of a phononic \Kxy~below
(and only below) the critical doping $p^{\star}$ would be evidence that spin correlations are a property
of the pseudogap phase.

\end{abstract}

\maketitle

\section{Introduction}

The discovery of a large negative thermal Hall conductivity~\Kxy~in hole-doped 
cuprates~\cite{Grissonnanche2019}, persisting down to zero doping ($p=0$), in the Mott insulating 
phase~\cite{Boulanger2020}, revealed a fascinating facet of cuprate high-$T_{\mathrm{c}}$ superconductors. 
This negative \Kxy, now attributed to phonons~\cite{Boulanger2020,Grissonnanche2020}, is a new experimental
signature of the enigmatic pseudogap phase of cuprates, as it disappears above the critical doping
$p^\star$ that marks the end of that phase.
However, the microscopic mechanism that makes phonons in cuprates acquire a handedness in a magnetic field
remains unknown.
A number of theoretical proposals have been made, including 
a coupling of phonons to magnons~\cite{ye2021} or spinons \cite{Samajdar2019,Zhangyc2021},
to collective fluctuations~\cite{mangeolle2022}, 
and to a state of loop-current order with the appropriate symmetries~\cite{Varma2020}, or 
the scattering of phonons by impurities or defects~\cite{Chen2020,Guo2021,Flebus2022,sun2022,Guo2022}.
It is still unclear which of these mechanisms, if any, is appropriate for cuprates.

A recent study of electron-doped cuprates~\cite{boulanger2022} showed that the negative thermal Hall
conductivity in cuprates is also present on the other side of the phase diagram. 
As a function of doping, the negative \Kxy~signal is observed from the insulating antiferromagnetic phase at
$x=0$ all the way up to $x = 0.17$, in the metallic phase above optimal doping. 
And here also, this negative \Kxy~was shown to be carried by phonons.

In the metallic phase, electrons are also expected to generate a thermal Hall signal due to
the Lorentz force.
In this study, we show that in our cleanest crystals of the electron-doped cuprate \ncco~(NCCO),
electrons and phonons make comparable contributions to the thermal Hall conductivity in the metallic state
at high doping, but with opposite sign.
The electronic contribution is positive, as expected from the Wiedemann-Franz law
and the known electrical Hall conductivity $\sigma_{\mathrm{xy}}$.
The negative phononic contribution is similar in magnitude and temperature dependence to that found
at lower doping~\cite{boulanger2022}.
This shows that the phonon thermal Hall conductivity of cuprates is independent of whether the host material
is an insulator or a metal.
This rules out any mechanism based on the skew scattering of phonons off charged impurities,
such as oxygen vacancies, as these local charges should be screened very effectively by mobile electrons
in a highly conductive metallic state.

To further characterize our samples, we sought to measure quantum oscillations.
The presence of low-frequency oscillations in our cleanest sample with $x=0.16$ shows that the Fermi surface is
reconstructed at that doping, clear evidence for the presence of significant antiferromagnetic correlations even at such high doping.
We speculate that these correlations may play a role in generating the phonon thermal Hall signal
in electron-doped cuprates, and possibly also in the pseudogap phase of hole-doped cuprates.

%-------------------------------------------------------------------------------------------------------------
\begin{figure}[t]
\centering
\includegraphics[width = \linewidth]{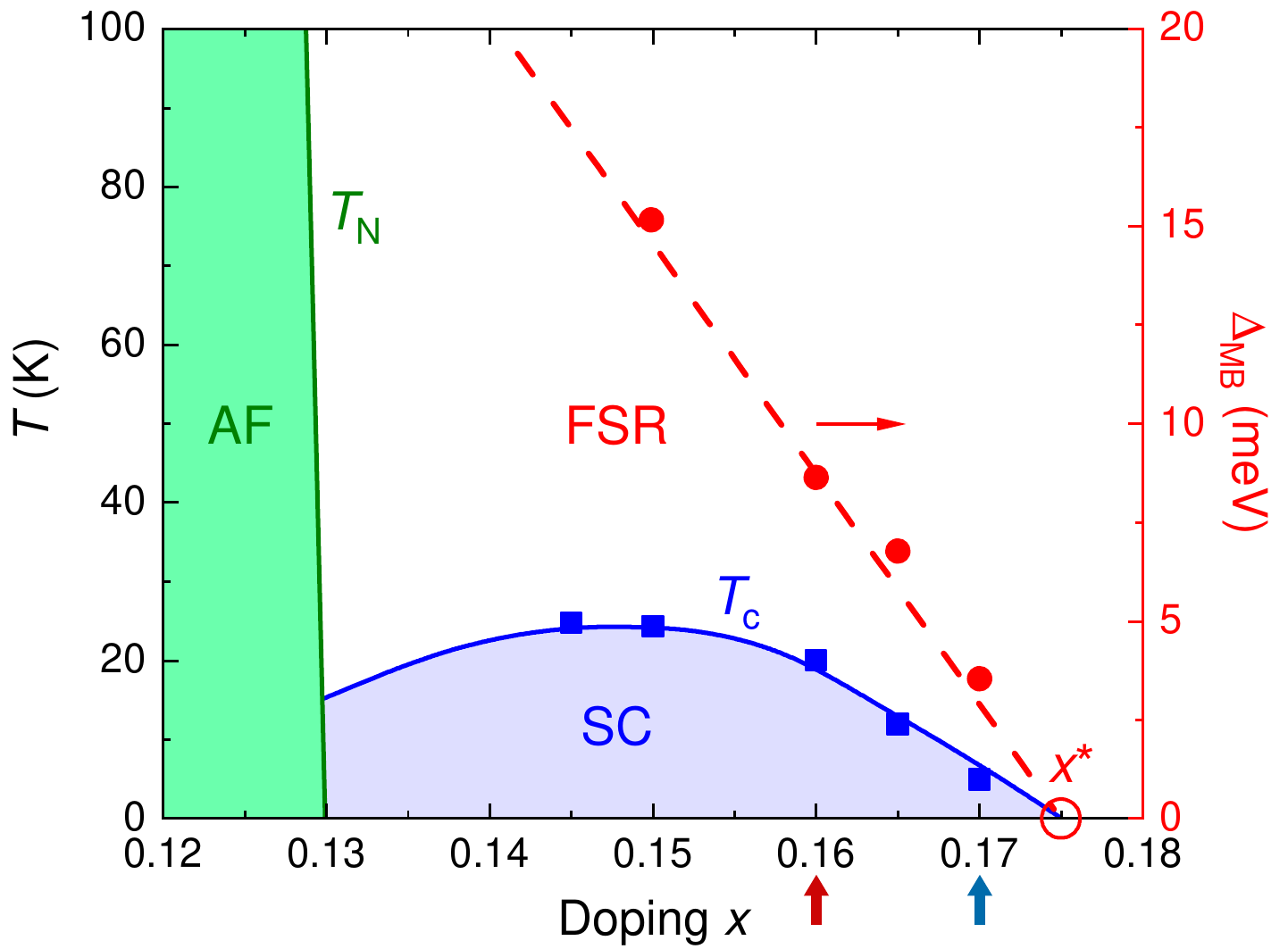}
\caption{
Temperature-doping phase diagram of the electron-doped cuprate NCCO. 
The antiferromagnetic phase (AF) is bounded by the N\'eel temperature \TN~(solid green line)
and the superconducting phase (SC) by the zero-field critical temperature \Tc~(blue squares) \cite{Helm2010}. 
The magnetic breakdown gap $\Delta_{\mathrm{MB}}$ (red circles), obtained in a prior study from quantum oscillations, 
is a measure of how strongly the Fermi surface of NCCO is reconstructed \cite{Helm2010}.
The doping dependence of $\Delta_{\mathrm{MB}}$ yields a critical doping \xstar~$= 0.175$ (open red circle)
above which the Fermi surface is not reconstructed.
The solid blue line and red dashed line are guides to the eye.
In our study, we investigated two samples of NCCO with nominal concentration
$x = 0.16$ (red arrow) and one with 
$x = 0.17$ (blue arrow).
}
\label{fig1}
\end{figure}
%-------------------------------------------------------------------------------------------------------------

\section{Methods}

\subsection{Samples}

Single crystals of \ncco~with nominal concentrations $x = 0.16$ (two samples) and $x = 0.17$ (one sample) were grown by the
traveling-solvent floating-zone method in O$_2$ and annealed in flowing argon for 48 hours at 900\,$^{\circ}$C. 
The three samples have a doping such that they lie in the superconducting/metallic regime of the phase diagram
(see Fig.~\ref{fig1}).
The superconducting transition temperature in zero field, defined by the onset of the drop in magnetization, 
is \Tc~$= 23.5$\,K and 20.0\,K for $x$ = 0.16 and 0.17, respectively.
For the transport measurements, crystals were cut into rectangular platelets with
dimensions (length between contacts $\times $ width $\times$ thickness, in $\mu$m) 
$620 \times 660 \times 90$ ($x = 0.16$, sample~1),
$1114 \times 1080 \times 185$ ($x = 0.16$, sample~2), and 
$1000 \times 1050 \times 80 $ ($x = 0.17$).
Contacts were made using silver epoxy, diffused at 500\,$^{\circ}$C under oxygen for 1 hour.
From the same mother crystals, another pair of NCCO samples were prepared, with $x = 0.16$ and $x = 0.17$,
for the tunnel diode oscillator measurements (see Section~\ref{sec:methods-TDO}). The dimensions of these samples (in $\mu$m) are
$730 \times 220 \times 100$ ($x= 0.16$) and 
$440 \times 300 \times 50$ ($x = 0.17$).

Note that it is difficult to achieve good enough electrical contacts on NCCO crystals to perform
a measurement of the in-plane electrical resistivity $\rho_{\mathrm{a}}$ that is free of $c$-axis current 
contamination in this highly 2D material.
For this reason, we have turned to measurements of the thermal conductivity \Kxx~down to very low
temperature (50\,mK) to access the residual resistivity $\rho_0$ of our samples
via the Wiedemann-Franz (WF) law, $\kappa_0/T = L_0/\rho_0$,
with $L_0 \equiv (\pi/3)(k_B/e)^2$.
Heat transport does not suffer from $c$-axis contamination.

\subsection{Thermal transport measurements}
The thermal conductivity \Kxx~is measured by applying a heat current $J_\mathrm{x}$ along the $x$ axis of the sample
(longest direction), which generates a longitudinal temperature difference
$\Delta T_{\mathrm{x}} = T^{+} - T^{-}$. 
The thermal conductivity \Kxx~is given by
\begin{align}
  \kappa_{\mathrm{xx}} = \frac{J_\mathrm{x}}{\Delta T_{\mathrm{x}}}\left(\frac{L}{wt}\right),
\end{align}
where $w$ is the sample width, $t$ its thickness, and $L$ the distance between $T^{+}$ and $T^{-}$.
When a magnetic field is applied perpendicular to the heat current ($H \perp J_\mathrm{x}$) (\ie~parallel to $z$), 
a transverse temperature difference $\Delta T_\mathrm{y}$ can develop along the $y$ axis. 
The thermal Hall conductivity \Kxy~is then given by 
\begin{align}
  \kappa_{\mathrm{xy}} = -\kappa_{\mathrm{yy}}\left(\frac{\Delta T_\mathrm{y}}{\Delta T_\mathrm{x}}\right)\left(\frac{L}{w}\right)
\end{align}
after antisymmetrization,
\ie~$\Delta T_\mathrm{y} (H) = \left[\Delta T_\mathrm{y} (T,+H) - \Delta T_\mathrm{y} (T,-H)\right]/2$.
Since NCCO is a tetragonal system, we can take $\kappa_{\mathrm{yy}} = \kappa_{\mathrm{xx}}$. 
The error bar on \Kxx~and \Kxy~is roughly $\pm 15 \%$ for each, coming mostly from the uncertainty
on sample dimensions and geometric factors.
The magnetic field was applied perpendicular to the CuO$_2$ planes.
A field of $H = 15$\,T is large enough to suppress superconductivity down to $T \to 0$~in all samples
\cite{Tafti2014}.

We use a steady-state method to measure both thermal conductivity \Kxx~and thermal Hall conductivity \Kxy. 
The data are taken while changing temperature in discrete steps at a fixed magnetic field.
The thermal gradient along the sample is provided by a resistive heater connected to one end of the sample. 
The other end of the sample is glued to a copper block with silver paint which acts as a heat sink.
Below 3\,K, the longitudinal temperature difference $\Delta T_\mathrm{x}$ is measured in a dilution refrigerator
down to 50\,mK, using two RuO$_2$ thermometers calibrated \textit{in situ} as a function of temperature
and magnetic field.
The transverse temperature difference $\Delta T_\mathrm{y}$ is measured using two RuO$_2$ thermometers 
connected to opposite sides of the sample.
Above 3\,K, $\Delta T_\mathrm{x}$ and $\Delta T_\mathrm{y}$ are measured in a standard variable temperature insert (VTI)
system up to 100\,K, using type-E thermocouples, known to have a weak magnetic field dependence.
More details of the technique can be found in
Refs.~\onlinecite{Boulanger2020,Grissonnanche2016,Grissonnanche2020,chen_planar_2024}.

\subsection{TDO measurements}
\label{sec:methods-TDO}
Quantum oscillations were detected using a tunnel diode oscillator (TDO) method in a magnetic field
up to 85\,T down to 1.8\,K. 
The sample is placed on a compensated spiral coil of a self-resonating LC circuit operating around 20~MHz. 
The circuit is driven by a tunnel diode polarized in its negative resistance region of the current-voltage
characteristic. 
The driving system (tunnel diode + tank capacitor) is separated from the TDO coil by a 1.25-meter-long coaxial
cable and the magnetic field was applied perpendicular to the CuO$_{2}$ planes.

In a TDO measurement, variations in the electrical resistance of the sample change the coil inductance
through skin depth \cite{Coffey2000} resulting in a shift in the TDO resonance frequency. 
By measuring this shift, one can detect Shubnikov-de Haas oscillations.

%-------------------------------------------------------------------------------------------------------------
\begin{figure}[t]
\centering
\includegraphics[width = 0.95\linewidth]{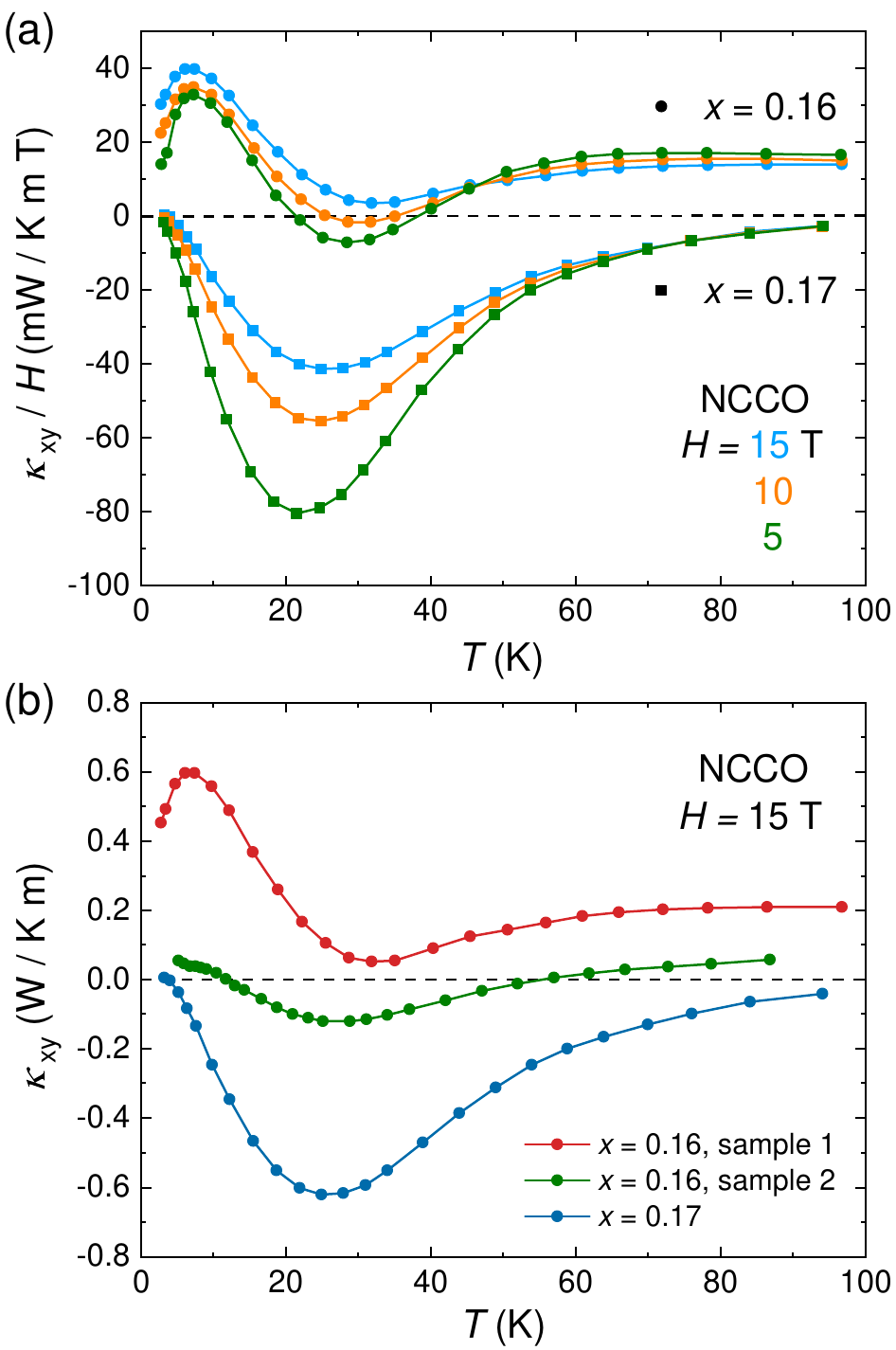}
\caption{
(a) Temperature dependence of the thermal Hall conductivity for NCCO $x = 0.16$, sample~1 (circles) and $x = 0.17$ (squares), plotted as $\kappa_{\mathrm{xy}} / H$ vs $T$ for fields $H = 5$\,T (green), 10\,T (orange), and 15\,T (blue). 
We observe two behaviors: 
the negative thermal Hall conductivity of phonons (very clear in the sample with $x = 0.17$ and also visible
in the sample with $x = 0.16$)
and the positive thermal Hall conductivity of electrons (visible only in the sample with $x=0.16$). 
(b) Thermal Hall conductivity of our three NCCO samples at $H = 15$\,T, plotted as $\kappa_{\mathrm{xy}}$
vs $T$. 
At $x=0.16$, sample~1 is cleaner than sample~2, resulting in a larger positive electronic contribution in the former.
}
\label{fig2}
\end{figure}
%-------------------------------------------------------------------------------------------------------------

\section{Results}

\subsection{Thermal transport}

In Fig.~\ref{fig2}(a), we display the thermal Hall conductivity of our two main samples, $x = 0.16$ (sample~1)
and $x=0.17$, plotted as $\kappa_{\mathrm{xy}} / H$ vs $T$ below 100\,K (down to 3\,K), for three different magnetic
fields $H$.
At $H=15$\,T, we find that \Kxy~is negative at all temperatures for the sample with $x=0.17$ (blue squares),
whereas it is positive at all temperatures for the sample with $x=0.16$.
The negative \Kxy~is similar to that found at lower doping~\cite{boulanger2022}, down to $x=0$, and it is
due to phonons.
The positive \Kxy~is due to electrons, as we show below.

\begin{figure}[t]
\centering
\includegraphics[width = 0.95\linewidth]{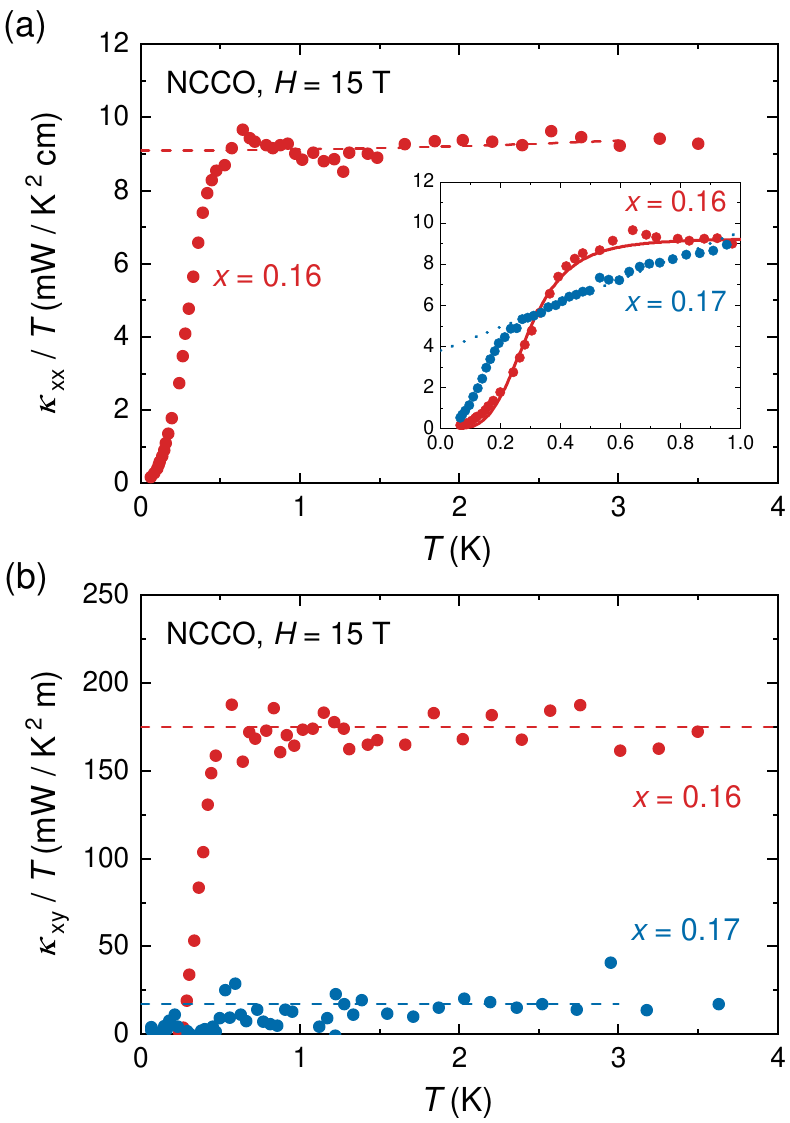}
\caption{(a) Thermal conductivity of NCCO $x=0.16$ (sample~1; red dots) at $H = 15$\,T, plotted as $\kappa_{\mathrm{xx}} / T$ vs $T$. 
The dashed line is a linear fit to extract the residual term in the form
$\kappa_{\mathrm{xx}}/T$$ = \beta T^2 + \kappa_0/T$, with
$\beta = 0.03$\,mW/K$^4$cm and
${\kappa^0_{\mathrm{xx}}}/{T} = 9.1 \pm 0.1$\,mW/K$^2$cm.
Inset: Thermal conductivity for NCCO $x=0.16$ (red) and $x = 0.17$ (blue) at $H = 15$\,T, plotted as
$\kappa_{\mathrm{xx}} / T$ vs $T$, below 1\,K.
In both samples, we observe a downturn in $\kappa_{\mathrm{xx}} / T$ as $T\to 0$, 
characteristic of electron-phonon decoupling \cite{Smith2005}. 
The red solid line is a fit of the $x = 0.16$ data
to Eq.~\ref{eq:decoup}, which gives a decoupling temperature of $T_{\mathrm{dec}}$ = 290\,mK.
The blue dotted line is a linear fit to the $x = 0.17$ data that gives a residual linear term of
${\kappa^0_{\mathrm{xx}}}/{T} = 4.0 \pm 0.1$\,mW/K$^2$cm for the $x = 0.17$ sample.
(b) Thermal Hall conductivity of NCCO for $x=0.16$ (sample~1) and $x = 0.17$ at $H = 15$\,T plotted as
$\kappa_{\mathrm{xy}} / T$ vs $T$.
We extract a residual linear term of $\kappa^0_{\mathrm{xy}}/T = 175 \pm 20$\,mW/K$^2$m for $x = 0.16$ and
$\kappa^0_{\mathrm{xy}}/T = 17.5 \pm 10$\,mW/K$^2$m for $x = 0.17$ (dashed lines).
}
\label{fig3}
\end{figure}

There is a contribution of phonons and electrons in both samples, but the electronic contribution is much larger in the sample with $x=0.16$.
This is confirmed by performing measurements down to very low temperature.
In Fig.~\ref{fig3}(b), we report our \Kxy~data at $H=15$\,T taken below 4\,K, plotted as \Kxy$/T$ vs $T$.
The fact that \Kxy$/T$ is constant below 4\,K shows that it is entirely coming from transport by fermions, 
\ie~electrons. Indeed, the contribution of phonons to \Kxy~in cuprates is negligible below
4\,K~\cite{Boulanger2020}.
We see that this (positive) electronic contribution is 10 times larger in the sample with $x=0.16$, 
as reflected in the residual linear term obtained by extrapolating \Kxy$/T$ to $T=0$
(dashed lines), namely $\kappa^0_{\mathrm{xy}}/T = 175 \pm 30$~mW/K$^2$m for $x = 0.16$ and 
$17.5 \pm 10$~mW/K$^2$m for $x = 0.17$ (Fig.~\ref{fig3}(b)).

The thermal Hall conductivity was also measured in NCCO $x=0.16$ sample~2, which is dirtier than sample~1. In Fig.~\ref{fig2}(b), we see that
the overall \Kxy~signal is more negative in sample~2 compared to sample~1, which indicates that
the positive electronic contribution is larger in a cleaner sample due to less scattering of electrons
by impurities.
%

%-------------------------------------------------------------------------------------------------------------
%-------------------------------------------------------------------------------------------------------------

%
We attribute the larger positive electronic contribution of \Kxy~ in $x = 0.16$ sample~1 to
the fact that this sample is cleaner than the $x = 0.17$ sample.
This can be verified by making a direct comparison of their thermal conductivity data as $T \to 0$. 
Figure \ref{fig3}(a) shows the longitudinal thermal conductivity, plotted as \Kxx/$T$ vs $T$,
at low temperature down to $T \to 0$. 
Since both samples are in the metallic region of the phase diagram, a residual linear term is expected. 
By applying a linear fit,
we obtain $\kappa_{0}/T = 9.1$\,mW/K$^2$cm for $x = 0.16$ and $\kappa_{0}/T = 4.0$\,mW/K$^2$cm for $x = 0.17$ (Fig.~\ref{fig3}(a), main panel, dashed line).
Using the Wiedemann-Franz (WF) law, we get $\rho_0 = 2.7 \pm 0.1\,\mu\Omega$\,cm for the former
and $\rho_0 = 6.1 \pm 0.1\,\mu\Omega$\,cm for the latter; there is roughly a factor of 2.3 between the two values of $\rho_0$, which clearly shows that the $x$ = 0.16 sample is cleaner than the $x$ = 0.17 sample.

As we can observe in Fig.~\ref{fig3}, there is a rapid downturn in both \Kxx~and \Kxy~below $T \sim 500$\,mK. 
A similar downturn was observed in Pr$_{2-x}$Ce$_x$CuO$_4$ (PCCO) at $T \approx 300$\,mK~\cite{Hill2001}, which was
attributed to the thermal decoupling between electrons and phonons at low temperature~\cite{Smith2005}. 
The decoupling temperature $T_{\mathrm{dec}}$ is extracted by fitting the $\kappa_{\mathrm{xx}}/T$ data to the expression
\begin{equation}
 \kappa_{\mathrm{xx}}/T = \alpha \frac{1}{1+\frac{r}{1+r(T/T_{\mathrm{dec}})^{n-1}}}+\beta T^{2},
 \label{eq:decoup}
\end{equation}
in which $\alpha$, $\beta$, $r$, $n$, and $T_{\mathrm{dec}}$ are all fitting parameters \cite{Smith2005}.
In the inset of Fig.~\ref{fig3}(a), the red solid line shows the thermal decoupling fit of our $x = 0.16$ data to Eq.~\ref{eq:decoup},
which gives a decoupling temperature of $T_{\mathrm{dec}} = 290$\,mK.

%-------------------------------------------------------------------------------------------------------------
\begin{figure}[t]
\centering
\includegraphics[width = 0.87\linewidth]{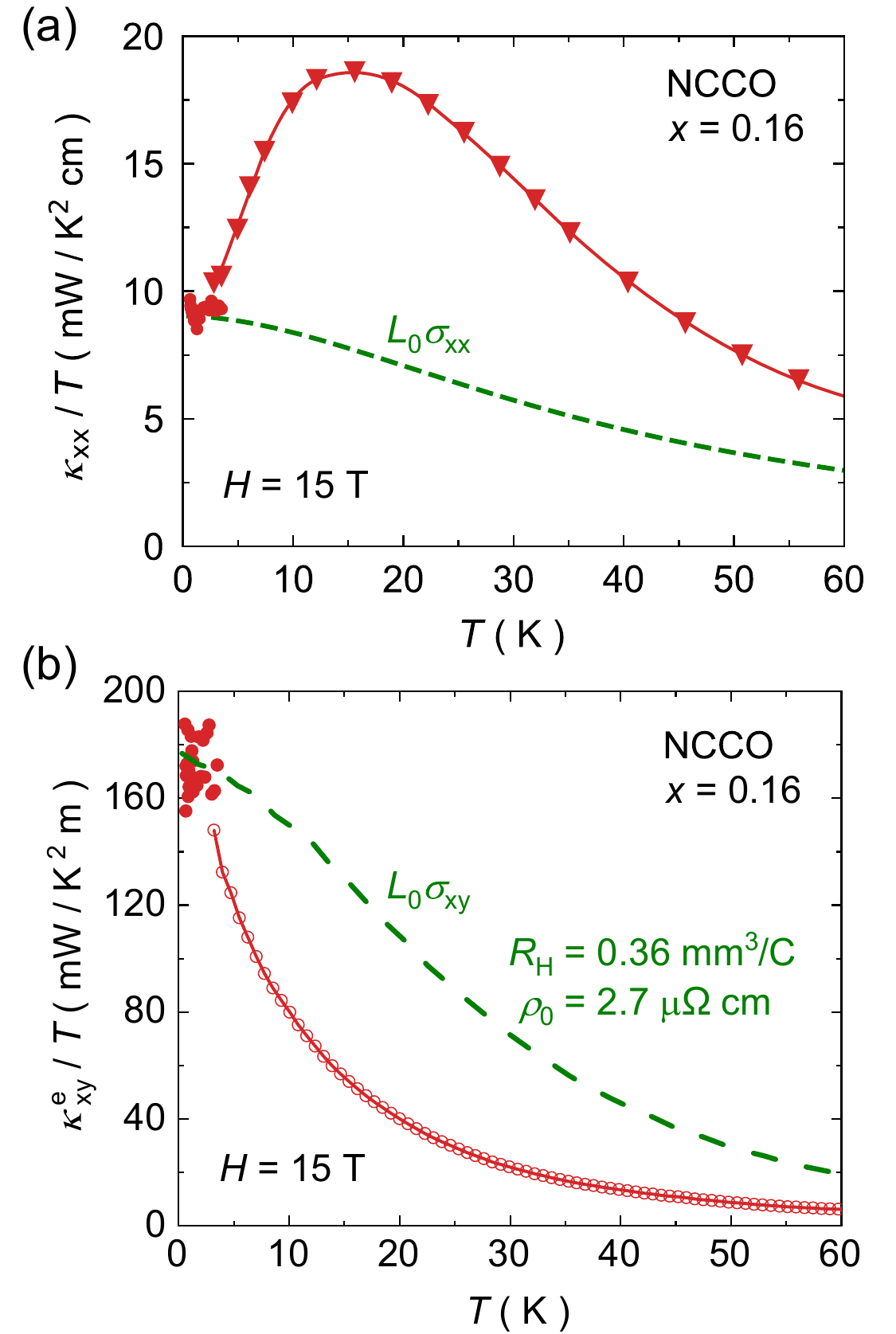}
\caption{(a)~Thermal conductivity of NCCO at $x = 0.16$ (sample 1, red) plotted as
$\kappa_{\mathrm{xx}}/T$ vs $T$ at $H = 15$\,T. 
The dashed green line is the electronic contribution obtained from the WF law (see text).
(b)~Electronic part of the thermal Hall conductivity plotted as $\kappa^{e}_{\mathrm{xy}}/ T$ vs $T$ for NCCO $x = 0.16$ (red) at $H = 15$\,T. 
For $T>4$\,K, $\kappa^{e}_{\mathrm{xy}}$ (open red circles) is obtained by subtracting the \Kxy~of the $x=0.17$ sample from the $x=0.16$ data (Fig.~\ref{fig2}(b)), assuming the phononic \Kxy~is the same in $x=0.16$ and $x=0.17$, and that the electronic contribution to the $x=0.17$ data is negligible.
The dashed green line represents an upper bound on the electronic contribution, set by
$L_{0}\sigma_{\mathrm{xy}}$ from the WF law. 
$L_{0}\sigma_{\mathrm{xy}}$ is calculated by using a temperature-independent Hall coefficient $R_\mathrm{H}$
and temperature-dependent $\rho_{\mathrm{xx}} (T)$ data of PCCO at $x = 0.17$~\cite{Tafti2014},
but adjusted for $\rho_0 = 2.7\,\mu\Omega$\,cm.
In order to match the value of $\kappa^{e}_{\mathrm{xy}}/T = 175 \pm 20$\,mW/K$^2$m at $T\rightarrow$ 0, we get a Hall
coefficient of $R_\mathrm{H}$ = 0.36\,mm$^{3}$/C, which is roughly half of the expected value (see text).}
\label{fig4}
\end{figure}
%-----------------------------------------------------------------------------------------------------------------------------

%
Figure~\ref{fig4}(a) shows the thermal conductivity of NCCO at $x = 0.16$ plotted as $\kappa_{\mathrm{xx}}/T$ vs $T$, at $H = 15$\,T for the full temperature range.
The data are a combination of the low temperature data (obtained in a dilution fridge, circles)
and the high temperature data (obtained in the VTI, triangles). 
In the same figure, we also plot an estimate of the maximal electronic contribution (dashed lines), 
calculated using the Wiedemann-Franz law: $\kappa_{\mathrm{xx}}^{e}/ T = L_0 \sigma_{\mathrm{xx}}$
with $\sigma_{\mathrm{xx}} = \rho_{\mathrm{xx}} / \left(\rho_{\mathrm{xx}}^2 + \rho_{\mathrm{xy}}^2\right)$.
For this estimate, we use the data of $\rho_{\mathrm{xx}} (T)$ in PCCO at $x = 0.17$ published in
Ref.~\onlinecite{Tafti2014}, but with the calculated value of $\rho_0$ obtained from our thermal conductivity data.
For $\rho_{\mathrm{xy}}(T)$, we use the data of PCCO at $x = 0.17$ published in
Ref.~\onlinecite{Charpentier2010}, with $\rho_{\mathrm{xy}}(T) \equiv R_\mathrm{H}(T) H$. 
Figure~\ref{fig4}(a) gives us an idea of the relative contribution of electrons and phonons to \Kxx.
The electronic part is equal to the green dashed line at $T \to 0$ and less than that at $T>0$. The phonon part is at least as large as the distance between the red curve and the green dashed line.

In Figure \ref{fig4}(b), we provide an estimate of the electronic thermal Hall conductivity, $\kappa^e_{xy}$,
at $x = 0.16$ (for sample 1), plotted as $\kappa^e_\mathrm{xy}/T$ vs $T$, for $H = 15$\,T. The red dots at
low temperature are the raw data in Fig.~\ref{fig3}(b), and they are essentially equal to
$\kappa^e_\mathrm{xy}/T$ since phonons contribute very little below 4\,K \cite{Boulanger2020}. For $T > 4$\,K,
our estimate of $\kappa^e_\mathrm{xy}/T$ is the red curve, obtained as the difference between the raw data for
$x = 0.16$ (sample 1; Fig.~\ref{fig2}(b), red dots) and the raw data for $x = 0.17$ (Fig.~\ref{fig2}(b),
blue dots). This assumes that $\kappa_\mathrm{xy}$ in the $x = 0.17$ sample is entirely phononic and that the
two samples have the same phonon contribution. The first assumption is reasonable given that
$\kappa^e_\mathrm{xy}$ is 10 times smaller in the $x = 0.17$ sample than in the $x = 0.16$ sample no.\ 1
(Fig.~\ref{fig3}(b)). The second assumption is not unreasonable given the data for the $x = 0.16$ sample no.\ 2
in Fig.~\ref{fig2}(b), which lie between the other two curves. This is what you would expect for a constant
negative phonon background to which an increasing positive electronic term is added.

Let us now compare our curve for $\kappa^e_\mathrm{xy}/T$ (red dots and curve in Fig.~\ref{fig4}(b)) to the
electrical Hall conductivity $\sigma_\mathrm{xy}$. In Fig.\ref{fig4}(b), we plot $L_0 \sigma_\mathrm{xy}$ vs
$T$ as the green dashed line, obtained from the definition
$\sigma_\mathrm{xy} = \rho_\mathrm{xy} / (\rho^2_\mathrm{xx} + \rho^2_\mathrm{xy})$. In the $T = 0$ limit,
the Wiedemann-Franz law tells us that $\kappa^e_\mathrm{xy}/T = L_0 \sigma_\mathrm{xy}$. So we impose that
equality, and thus obtain the value of $\rho_\mathrm{xy} = R_\mathrm{H} H$ at $T \to 0$, given that we know
the value of $\rho_\mathrm{xx}$ at $T \to 0$, namely $\rho_0 = 2.7\,\mu$m\,cm. We get a Hall coefficient equal
to $R_\mathrm{H} = 0.36$\,mm$^3$/C. This is half the value that one must get for an unreconstructed Fermi
surface, namely $R_\mathrm{H} = V / (n e) = 0.7$\,mm$^3$/C, where $V$ is the volume per Cu atom, $e$ is the
electron charge and $n = 1-x$ is the carrier density contained in the full Fermi surface at a doping $x$. The
fact that we get a Hall coefficient that is lower than that reference value is consistent with the Fermi
surface being reconstructed, \ie~with $x < x^{\star}$ \cite{Charpentier2010}; such Fermi-surface reconstruction in NCCO up to $x^{\star}$ has been confirmed by recent ARPES work \cite{He2019}.

%-------------------------------------------------------------------------------------------------------------
\begin{figure}[t]
\centering
\includegraphics[width = 0.95\linewidth]{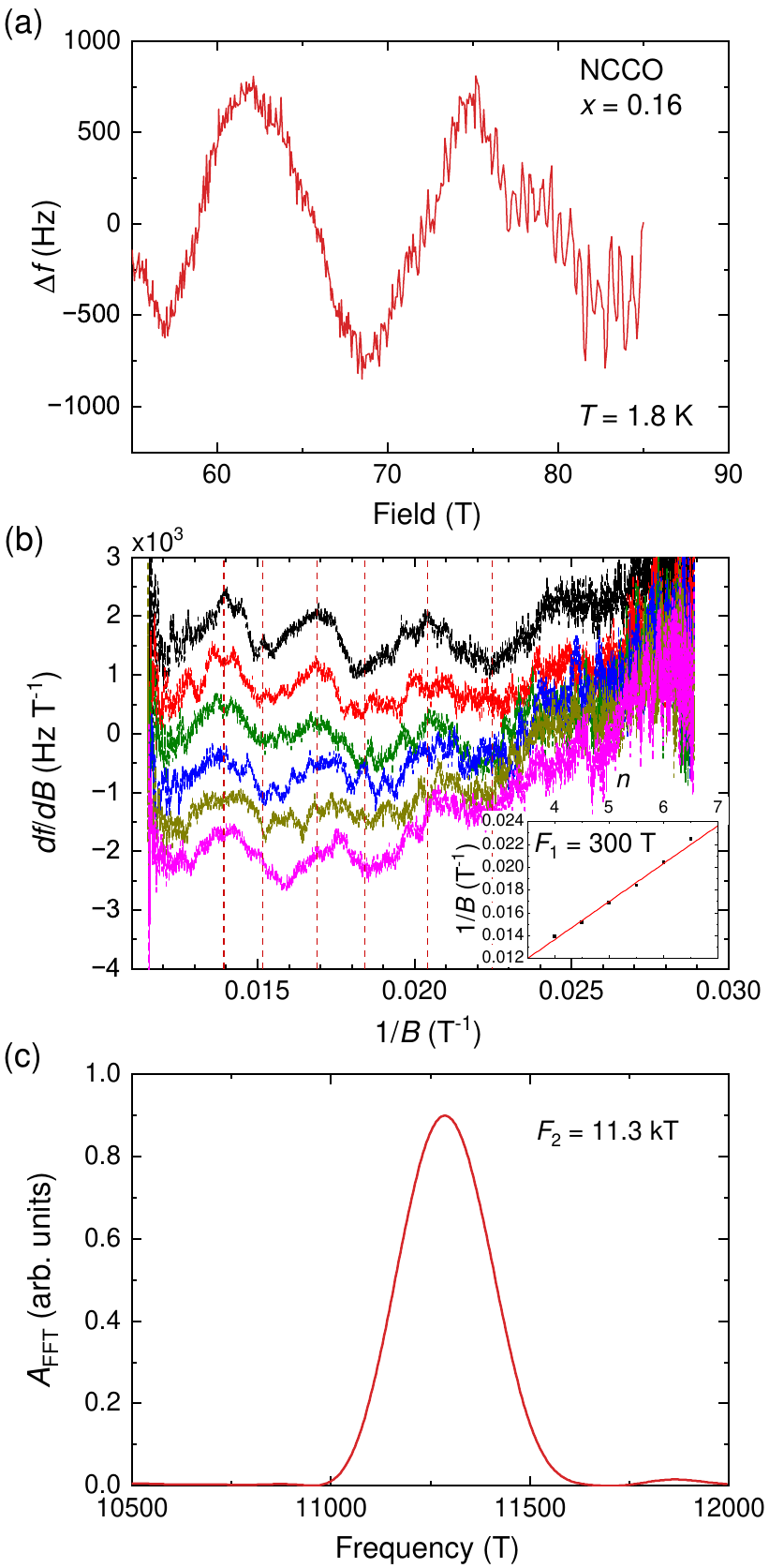}
\caption{Quantum oscillation pattern observed by TDO measurement. 
(a) The oscillatory part of the TDO signal after removing a smooth background at $T = 1.8$\,K.
(b) Plot of TDO signal derivative $df/dB$ vs. $1/B$ with locations of extrema marked; traces vertically offset for visibility. Inset: $1/B$ vs. index $n$ of extrema; slope of linear fit yields $F_1 = 300 \pm 20$~T.
(c) FFT of data showing the high-frequency peak at $F_2 = 11 300 \pm 100$~T.
}
\label{fig5}
\end{figure}
%-------------------------------------------------------------------------------------------------------------

\subsection{Quantum oscillations}

Separate evidence that the NCCO $x = 0.16$ sample~1 is cleaner than the $x = 0.17$ sample comes from our
quantum oscillation data. 
By performing TDO measurements in identical conditions for both $x = 0.16$ (sample~1) and $x = 0.17$ up to 85\,T,
we were able to observe quantum oscillations in the $x = 0.16$ sample but not in the $x = 0.17$ sample,
which suggests that the $x = 0.16$ sample is indeed cleaner than the $x = 0.17$ sample, since the amplitude of
quantum oscillations is exponentially proportional to the electronic mean free path.

The quantum oscillation data taken in the $x = 0.16$ sample (after subtracting a smooth background) 
are shown in Fig.~\ref{fig5}(a). Two oscillatory components are visible: a low frequency with $\sim{}2$ periods
between 50\,T and 90\,T, and a high frequency clearly visible above 80\,T.
A determination of the lower frequency is made by locating and indexing the peaks and troughs in the derivative $df/dB$ (Fig.~\ref{fig5}(b)), with the slope of a linear fit to $1/B$ vs. $n$ (inset) yielding $F_1 = 300 \pm 20$~T.
The high-frequency oscillations are readily separable from signal background, and a discrete Fourier transform suffices for a reliable determination of their frequency $F_2 = 11300 \pm 100$~T (Fig.~\ref{fig5}(c)). 

Below $x^{\star} = 0.175$ (Fig.~\ref{fig1}), the Fermi surface of NCCO undergoes a reconstruction due to the antiferromagnetic ordering,
which results in small electron-like and hole-like pockets \cite{Lin2005}.
The low frequency oscillations ($F_1$) come from the small hole pocket, while the high frequency ($F_2$)
can be explained by either a remnant ``gossamer'' large Fermi surface~\cite{Xu_2023} or by the magnetic breakdown between the hole- and electron-like pockets of the reconstructed Fermi surface~\cite{Helm2010, Helm2013}. 
In either case, this high frequency $F_2$ effectively corresponds to the area of an orbit around the large unreconstructed Fermi surface. Its value is given by
$F = n\Phi_0$, where the 2D carrier density is equal to $n = 1-x$ per Cu. Given $F = 11.3$\,kT, 
we get $x = 0.160$ if we assume that the in-plane lattice spacing has a value of 3.92~\AA. This value is within the range
of values quoted in the literature, namely $3.91 \pm 0.01$~\AA{} \cite{Schultz1996}
and $3.95 \pm 0.01$~\AA{} \cite{helm2009}.
This frequency tells us that the average radius of the roughly circular Fermi surface is given by the Onsager relation $F = \hbar k_F^2/2e$,
from which we infer $k_F = 0.586 \pm 0.003$~\AA{}$^{-1}$.

%-------------------------------------------------------------------------------------------------------------
\begin{figure}[t]
\centering
\includegraphics[width = 0.95 \linewidth]{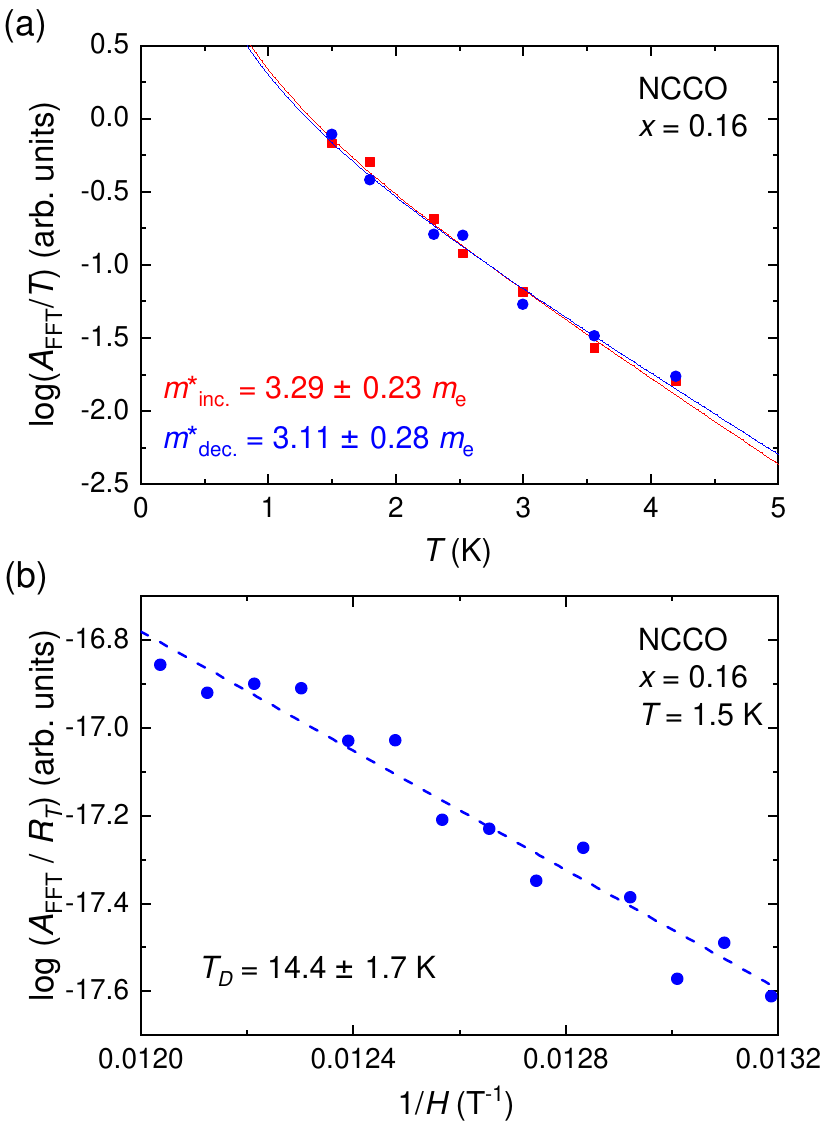}
\caption{
(a)~The temperature dependence of the FFT amplitude for the high frequency oscillation ($F_2$), normalized in the $T\to 0$ limit.
Separate fits to Eq.~\ref{eq:RThermal} (solid lines) for data collected during the increasing (red) and decreasing (blue) stages of the field pulse yield effective masses of $m_\mathrm{inc}^* = 3.29\pm0.23~m_e$ and $m_\mathrm{dec}^* = 3.11\pm0.28~m_e$, respectively; we use $m^* = 3.20 \pm 0.18~m_e$ for subsequent analysis.
(b)~Dingle plot of the amplitude vs magnetic field for the same oscillation at $T = 1.5$\,K,
giving a Dingle temperature $T_D = 14.4 \pm 1.7$~K. 
The dashed line is a fit to Eq.~\ref{eq:RDingle}.
}
\label{fig6}
\end{figure}
%-------------------------------------------------------------------------------------------------------------

From the Lifshitz-Kosevich (LK) theory, the temperature and magnetic field dependence of the oscillation
amplitude are determined by the product of the thermal damping factor $R_{T}$ and the Dingle damping
factor $R_{D}$, defined as
\begin{align}
  R_{T} &= \frac{\alpha T m^*/m_e}{B\sinh\left(\alpha T m^*/m_e B\right)},\label{eq:RThermal} \\
  R_{D} &= \exp\left[-\alpha T_{D}m^*/m_e B\right],\label{eq:RDingle}
\end{align}
where the Dingle temperature $T_{D} = \hbar / 2\pi k_{\mathrm{B}}\tau$,
the constant $\alpha \equiv 2\pi^2k_{\mathrm{B}} m_e / e \hbar = 14.69$\,T/K,
$\tau$ is the scattering time, 
$k_{\mathrm{B}}$ is the Boltzmann constant, 
$m^*$ is the effective carrier mass,
and $m_e$ is the bare electron mass.
Equation~\ref{eq:RDingle} can also be expressed in terms of the electronic mean free path $\ell_e$:
$R_D = \exp\left[-\pi r_c/\ell_e\right]$, where the cyclotron radius $r_c = m^*v_F/eB$.
We can then fit the temperature dependence of the normalized damping factor (Fig.~\ref{fig6}(a)) to Eq.~\ref{eq:RThermal}
to obtain an effective mass of $m^* = 3.20 \pm 0.18~m_e$ for the high frequency;
this allows us to extract the Fermi velocity $v_F = \hbar k_F/m^* = (2.12 \pm 0.12) \times 10^5$\,m/s.
Figure~\ref{fig6}(b) shows the field dependence of the oscillation amplitude. A fit to Eq.~\ref{eq:RDingle} yields a Dingle
temperature of $T_D = 14.4 \pm 1.7$~K, corresponding to a scattering lifetime $\tau = (8.4 \pm 1.0) \times 10^{-14}$\,s
and a mean free path $\ell_e = v_F \tau = 178 \pm 14$~\AA{}.
Both the effective mass and the Dingle temperature extracted here are consistent with previous quantum oscillation results
on NCCO at $x =$ 0.15 \cite{helm2009, Helm2013}.

\section{Discussion}

The main finding of this study is the positive thermal Hall conductivity of our NCCO $x = 0.16$ sample no. 1, compared to a large negative \Kxy~signal in the $x = 0.17$ sample (Fig.~\ref{fig2}(b)). 
Knowing from a previous study~\cite{boulanger2022} that the \Kxy~observed at $x = 0.17$ is dominated
by phonons, we propose two distinct contributions to \Kxy~in NCCO at high doping ($x > 0.15$):
a positive contribution from electrons, and a negative contribution from phonons. 
That the electronic \Kxy~is larger in the $x=0.16$ sample is simply due to the fact that the $x=0.16$ sample is cleaner (\ie~it has a longer electronic mean free path)
than the $x=0.17$ sample, as shown by the smaller residual resistivity $\rho_{0}$
of the $x=0.16$ sample and the absence of quantum oscillations in the $x=0.17$ sample.
At fixed nominal doping ($x = 0.16$), samples of different quality exhibit different magnitude of $\kappa^e_\mathrm{xy}$
(Fig.~\ref{fig2}(b)).

Combining these results with a previous study \cite{boulanger2022}, we find that the large negative phonon
contribution has similar magnitude and temperature dependence at all doping levels of the electron-doped cuprates NCCO and PCCO.
This observation indicates that the mechanism behind the phonon thermal Hall effect in electron-doped cuprates
is similar for all doping levels, regardless of whether the system is an insulator or metal.
This places strong a constraint on possible mechanisms behind the phonon thermal Hall effect and rules out any
mechanism based on the skew scattering of phonons off charged impurities \cite{Flebus2022}, such as
oxygen vacancies, as these local charges should be screened very effectively by mobile electrons
in a highly conductive metallic state.

So what makes phonons chiral in cuprates?
One possible explanation is the scattering of phonons by impurities embedded within an antiferromagnetic environment
\cite{Guo2022}.
Both the magnitude of the extracted Hall coefficient ($R_\mathrm{H}<0.7$\,mm$^{3}$/C) and the
presence of a low frequency in the quantum oscillation data in our NCCO $x=0.16$ sample confirm its proximity to 
the Fermi surface reconstruction point $x^*$, with $x < x^*$, where short-range antiferromagnetic correlations are still prominent. 
A large negative phonon contribution, similar in magnitude and temperature dependence, persists from the
antiferromagnetically ordered Mott insulating state all the way up to the metallic state close to $x^{\star}$,
which is consistent with a scenario of antiferromagnetic ordering playing a key role in the phonon thermal Hall effect.

It is worth noting that a similar \Kxy~consisting of both a negative phononic contribution and a positive
electronic contribution has also been observed in the hole-doped cuprate \ndlsco~ at $p$ = 0.20 \cite{Grissonnanche2019}, and high-field NMR measurements of the closely related La$_{2-x}$Sr$_x$CuO$_4$ have revealed the presence of an antiferromagnetic spin-glass phase from $p = 0.02$ all the way to $p^* \approx 0.19$ \cite{Frachet2020}.
The similar behavior of the phonon thermal Hall effect in hole-doped cuprates and electron-doped cuprates
indicates that the same mechanism involving antiferromagnetic correlations could potentially also explain the
phonon \Kxy~in hole-doped cuprates, as has been suggested by a recent study of impurity-induced
phonon thermal Hall effect in the antiferromagnetic phase of Sr$_{2}$IrO$_{4}$~\cite{Ataei2024}.

\section{Summary}

We have measured the thermal conductivity \Kxx~and the thermal Hall conductivity \Kxy~of the electron-doped 
cuprate NCCO at dopings $x = 0.16$ and $x = 0.17$. 
In the $x = 0.16$ sample, we observe two distinct channels for thermal Hall conductivity
in the metallic state:
a positive channel contributed by electrons and a negative channel contributed by phonons.
The fact that the negative phononic contribution is comparable in magnitude and temperature dependence 
regardless of the system being an insulator (at low doping) or a metal (at high doping)
indicates that its origin is unlikely to come from skew scattering of phonons off charged
impurities, since such a mechanism should depend strongly on the screening from mobile electrons.
The fact that the negative phonon thermal Hall effect persists from the antiferromagnetically ordered phase all the way
up to $x^{\star}$ (Fig.~\ref{fig1}), where short-range antiferromagnetic correlations are still present, is consistent with 
spin texture playing a role in the underlying mechanism for the phonon thermal Hall effect.
This type of mechanism, whereby phonons are scattered by spin texture or defects embedded in a magnetic
order, could also apply to hole-doped cuprates inside their pseudogap phase (below the critical doping
$p^{\star}$).

\section{Acknowledgments}

We thank S.~Fortier for his assistance with the experiments, and the cryogenics team at the Institut Quantique
for their support.
L.\,T. acknowledges support from the Canadian Institute for Advanced Research (CIFAR) as a CIFAR Fellow, 
and funding from the Institut Quantique,
the Natural Sciences and Engineering Research Council of Canada (Grant No.~PIN:123817),
and a Canada Research Chair.
C.\,P. acknowledges support
from the EUR grant NanoX No.~ANR-17-EURE-0009 and
from the ANR grant NEPTUN No.~ANR-19-CE30-0019-01.
G.\,G. acknowledges support from STeP2 No.~ANR-22-EXES-0013,
QuantExt No.~ANR-23-CE30-0001-01,
Audace CEA No.~ANR-24-RRII-0004,
and the \'{E}cole Polytechnique Foundation.
Z.\nobreakdash-X.\,S. acknowledges the support of the U.S. Department of Energy, Office of Science, Office of Basic Energy Sciences, Division of Material Sciences and Engineering, under contract DE-AC02-76SF00515.
M.\nobreakdash-E.\,B., L.\,C., G.\,G., J.\,B., and L.\,T. have benefited from their affiliation to the RQMP~\cite{RQMP}.
This research was undertaken thanks, in part, to funding from the Canada First Research Excellence Fund.
This research project No.~324046 is made possible thanks to funding from the Fonds
de recherche du Qu\'{e}bec.

\vfill

\bibliography{reference}

\end{document}